\documentclass[iop]{emulateapj}

\newcommand{\Ms}{M$_\odot$}

\slugcomment{Accepted by ApJL}
\shorttitle{VANDAM Class 0/I Candidate Disks}
\shortauthors{Segura-Cox et al.}

\begin{document}


\title{The VLA Nascent Disk and Multiplicity Survey: First Look at
Resolved Candidate Disks around Class 0 and I Protostars in the Perseus Molecular Cloud}

\author{Dominique M. Segura-Cox\altaffilmark{1}, Robert J. Harris\altaffilmark{1},  John J. Tobin\altaffilmark{2}, Leslie W. Looney\altaffilmark{1}, Zhi-Yun Li\altaffilmark{3}, Claire Chandler\altaffilmark{4}, Kaitlin Kratter\altaffilmark{5}, Michael M. Dunham\altaffilmark{6}, Sarah Sadavoy\altaffilmark{7}, Laura Perez\altaffilmark{4}, Carl Melis\altaffilmark{8}}
\altaffiltext{1}{Department of Astronomy, University of Illinois, Urbana, IL 61801, USA; segurac2@illinois.edu}
\altaffiltext{2}{Leiden Observatory, Leiden University, P.O. Box 9513, 2000-RA Leiden, The Netherlands}
\altaffiltext{3}{Department of Astronomy, University of Virginia, Charlottesville, VA 22903, USA}
\altaffiltext{4}{National Radio Astronomy Observatory, Socorro, NM 87801, USA}
\altaffiltext{5}{Steward Observatory, University of Arizona, Tucson, AZ 85721, USA}
\altaffiltext{6}{Harvard-Smithsonian Center for Astrophysics, Cambridge, MA 02138, USA}
\altaffiltext{7}{Max-Planck-Institut f{\"u}r Astronomie, D-69117 Heidelberg, Germany}
\altaffiltext{8}{Center for Astrophysics and Space Sciences, University of California, San Diego, CA 92093, USA}

\begin{abstract}
We present the first dust emission results toward a sample of seven protostellar disk candidates around Class 0 and I sources in the Perseus molecular cloud from the VLA Nascent Disk and Multiplicity (VANDAM) survey with $\sim$0.05$^{\prime\prime}$ or 12 AU resolution. 
 To examine the surface brightness profiles of these sources, we fit the Ka-band 8 mm dust-continuum data in the \textit{u,v}-plane to a simple, parametrized model based on the Shakura-Sunyaev disk model. The candidate disks are well-fit by a model with a disk-shaped profile and have masses consistent with known Class 0 and I disks.  The inner-disk surface densities of the VANDAM candidate disks have shallower density profiles compared to disks around more evolved Class II systems.  The best-fit model radii of the seven early-result candidate disks are $R_{c}$ $>$ 10 AU; at 8 mm, the radii reflect lower limits on the disk size since dust continuum emission is tied to grain size and large grains radially drift inwards. These relatively large disks, if confirmed kinematically, are inconsistent with theoretical models where the disk size is limited by strong magnetic braking to $<$ 10 AU at early times. 
\end{abstract}

\section{Introduction}

Disks around young protostars are intrinsically linked to planet formation, binary formation, and protostellar mass 
accretion \citep{wc11,ar11}.  
In spite of their importance in star formation, when and how disks form--and their properties at early times--are poorly constrained. Hence, whether large and massive disks can form in young protostars \citep[i.e.~Class 0/I stages,][]{an00,du14} is subject to debate \citep[e.g.,][respectively]{jg09,ch08,ma10}.  Theoretical studies demonstrate magnetic fields can affect the formation timescales and properties of disks.  In particular, strong magnetic fields can remove angular momentum from a collapsing envelope via magnetic braking, reducing forming disks to $R$ $<$ 10 AU \citep[e.g.,][]{ml08,db10,ma11,li11,dbk12}.  
Complicating the issue, several mechanisms can lessen the effects of magnetic braking, leading to larger disks: misalignment between the rotation axis and magnetic field of the system \citep[e.g.,][]{jo12,li13}, turbulence \citep[e.g.,][]{se13,jo13}, and initial conditions \citep[e.g.,][]{ma14}.
The protostellar stage represents the stage with the largest mass reservoir available to form disks, therefore understanding the properties of disks at early epochs is crucial to determine the formation mechanism behind these structures.

Protostellar disks, however, are difficult to observe.
Class 0 protostars and disks are so enshrouded that $\gtrsim$90\% of their  
millimeter emission comes from
the envelope \citep{lo00}.  
The three Class 0 protostars that have been observed with 
enough resolution and sensitivity to determine their disk properties and detect Keplerian rotation (L1527, VLA 1623, and HH212) have $R$ $>$ 30 AU disks, larger than expected 
from strong magnetic braking models \citep{oh14,to12,mu13,cd14}.
Observational limitations prevent smaller disks from being detected; these Class 0 disks may not represent typical disks at this stage of evolution.
Class I protostars are less embedded and have cleared enough of their mass reservoir that more disks have been detected than in Class 0 systems \citep[e.g.,][]{ha14}, though not as many disks have been revealed as in more-evolved Class II sources \citep[e.g.,][]{an09,an10}.
Without detection and study of more disks around 
 the youngest protostellar systems, we cannot characterize early 
conditions of mass accretion onto the central protostar and planet formation because evolutionary mechanisms can change disk properties by the Class II phase \citep{wc11}.

 \begin{figure*}[t]
        \centering
                \includegraphics[width=0.495\textwidth]{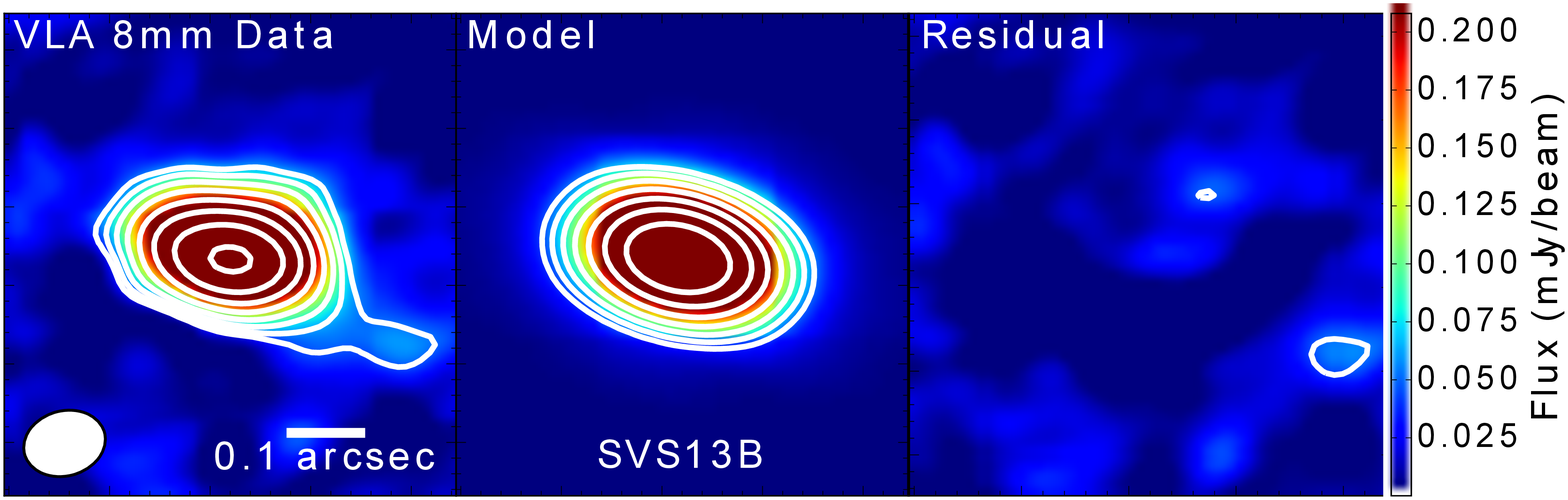}
                \label{fig1a}
                \includegraphics[width=0.495\textwidth]{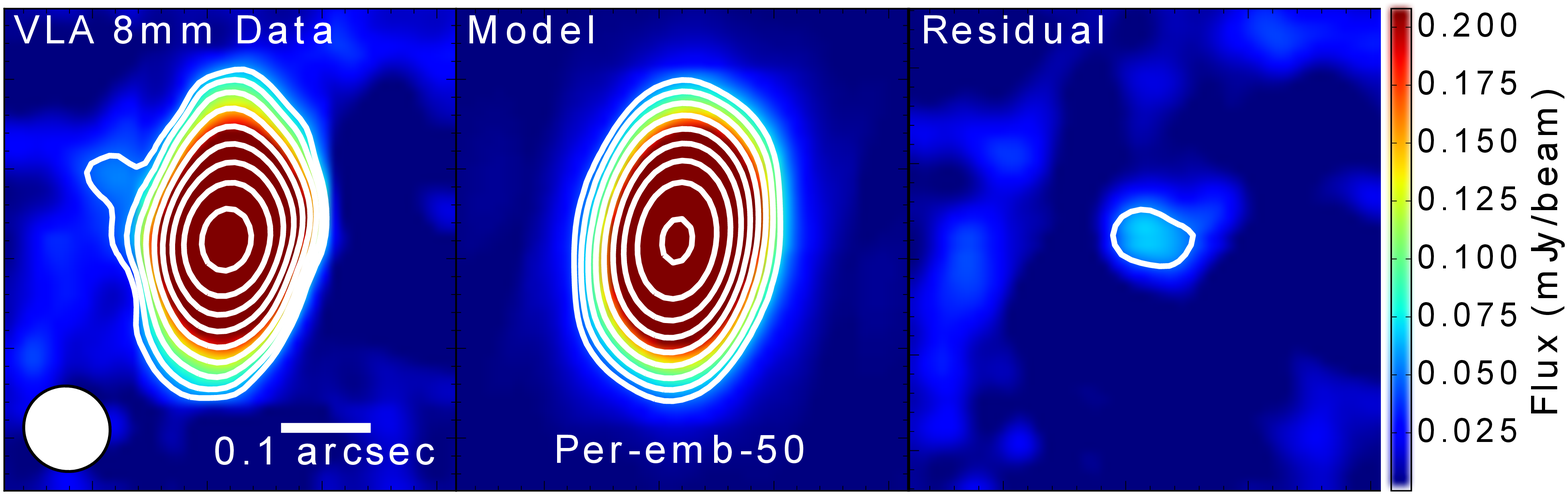}
                \label{fig1b}
                \includegraphics[width=0.495\textwidth]{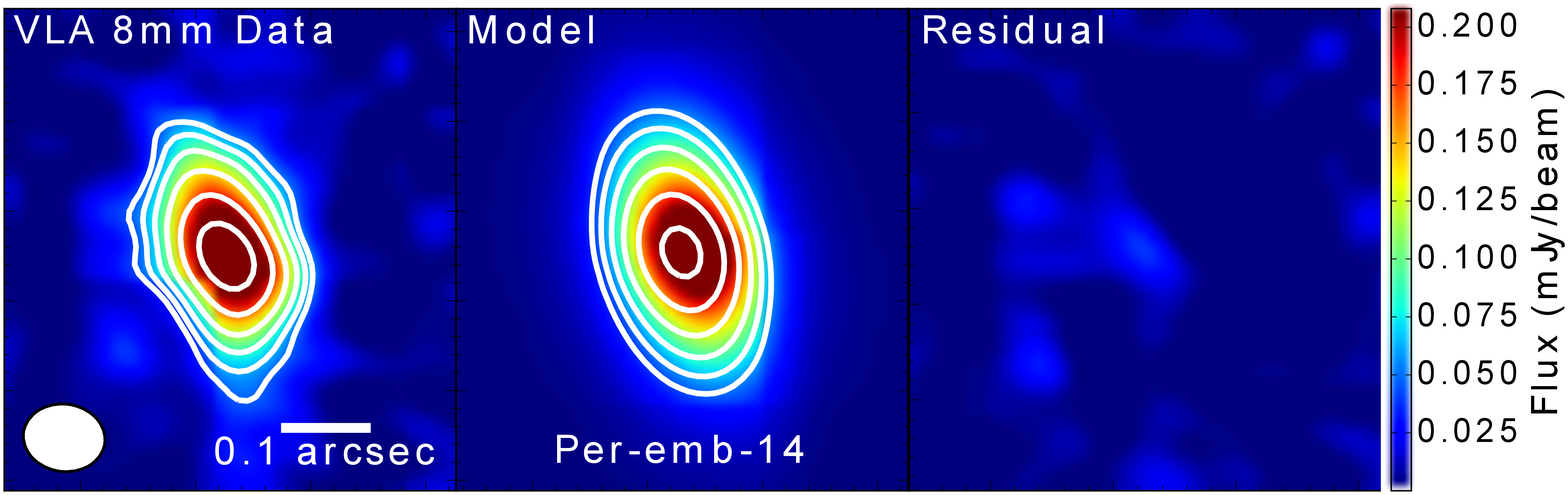}
                \label{fig1c} 
                \includegraphics[width=0.495\textwidth]{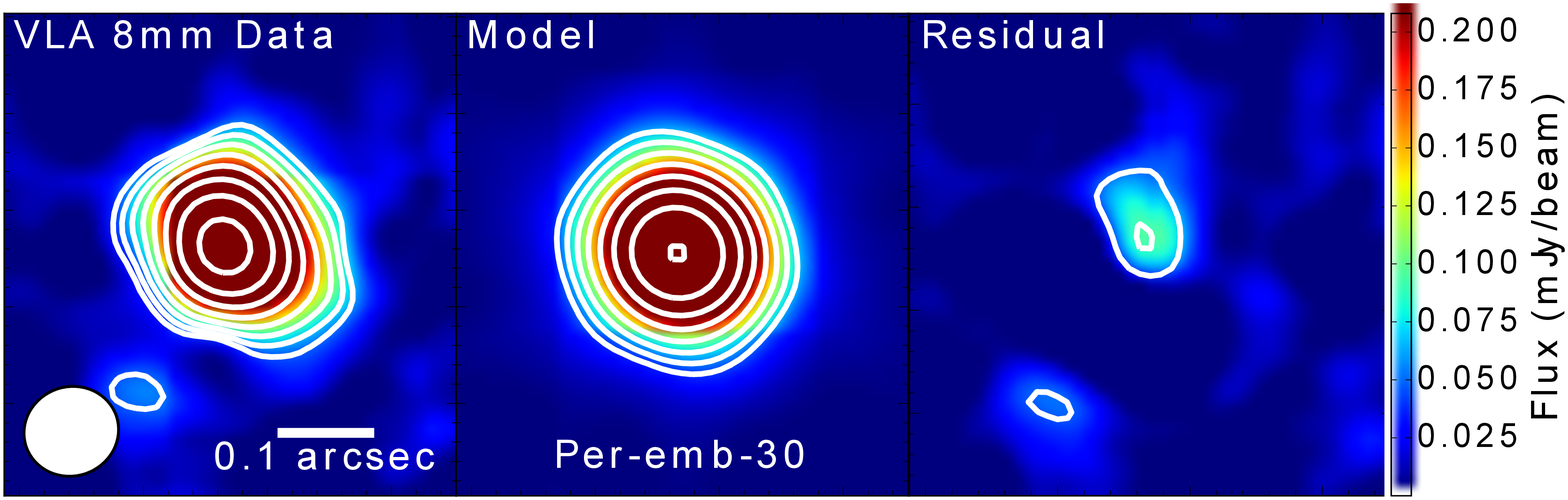}
                \label{fig1d} 
                \includegraphics[width=0.495\textwidth]{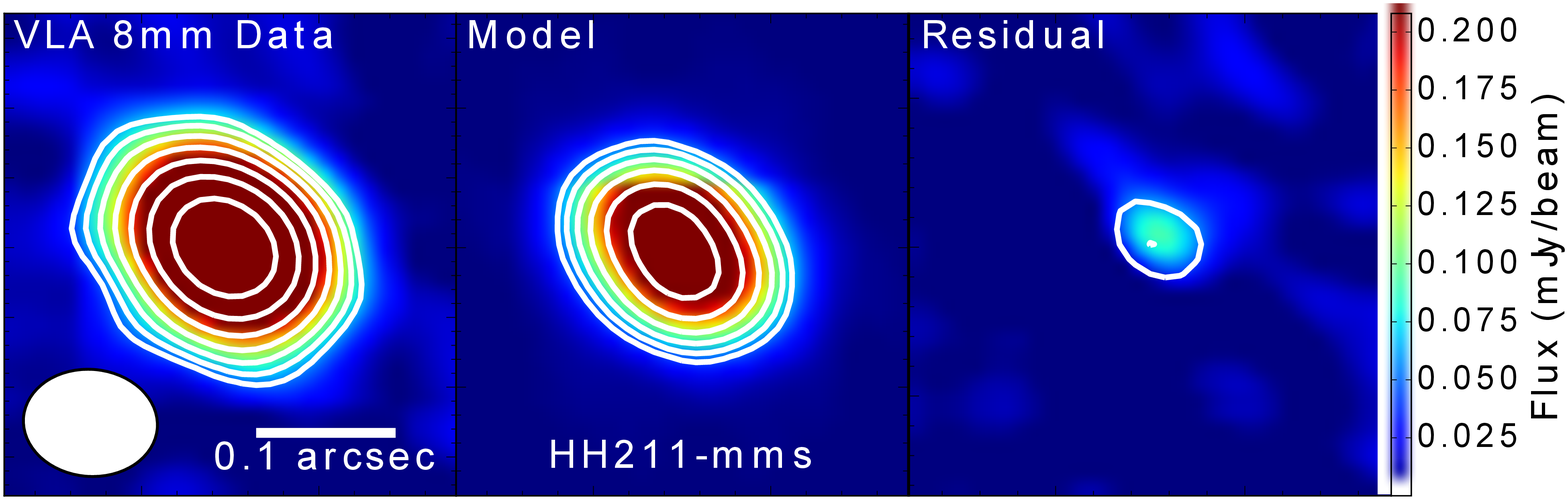}
                \label{fig1e}
                \includegraphics[width=0.495\textwidth]{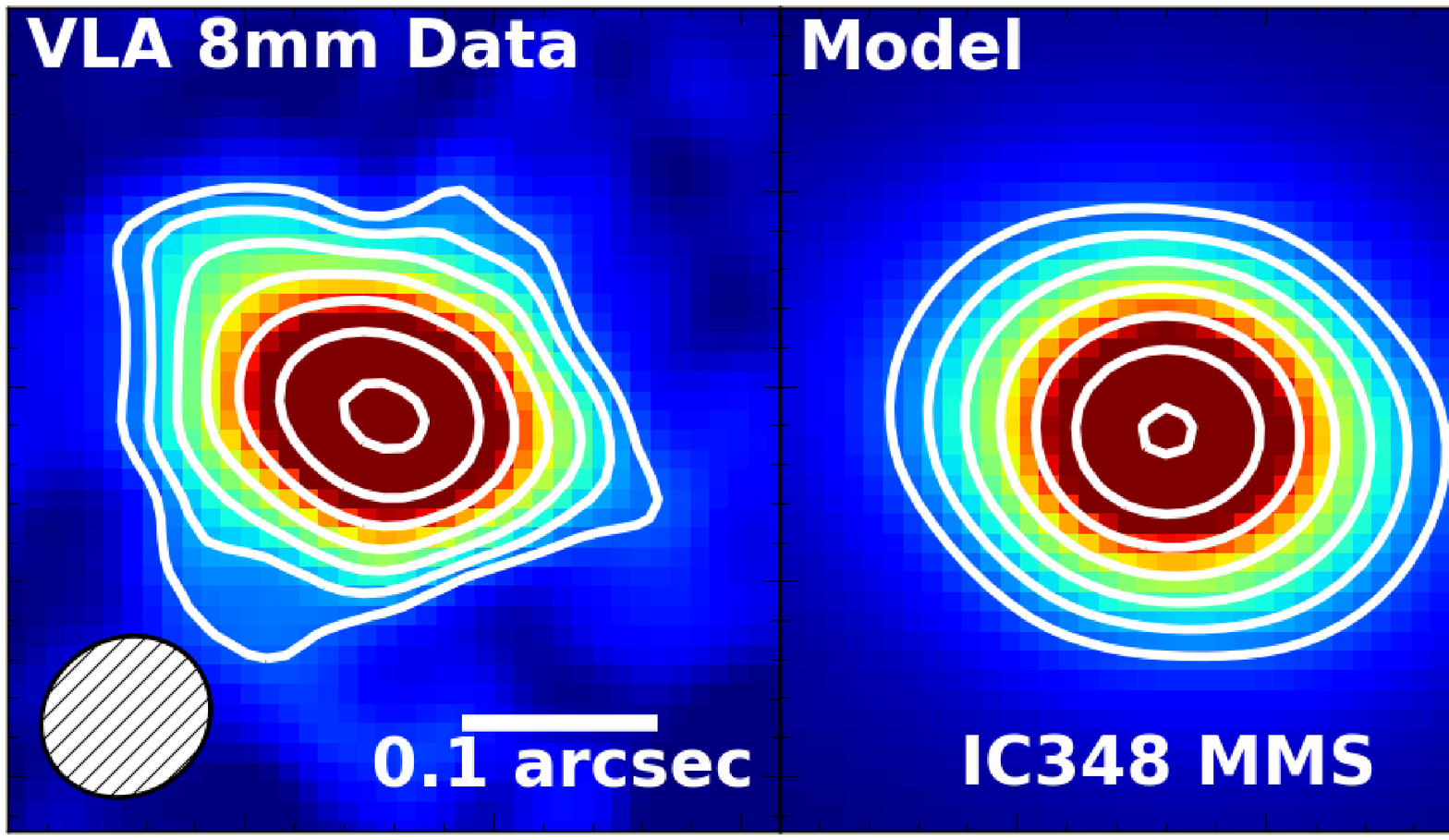}
                \label{fig1f}
                \includegraphics[width=0.495\textwidth]{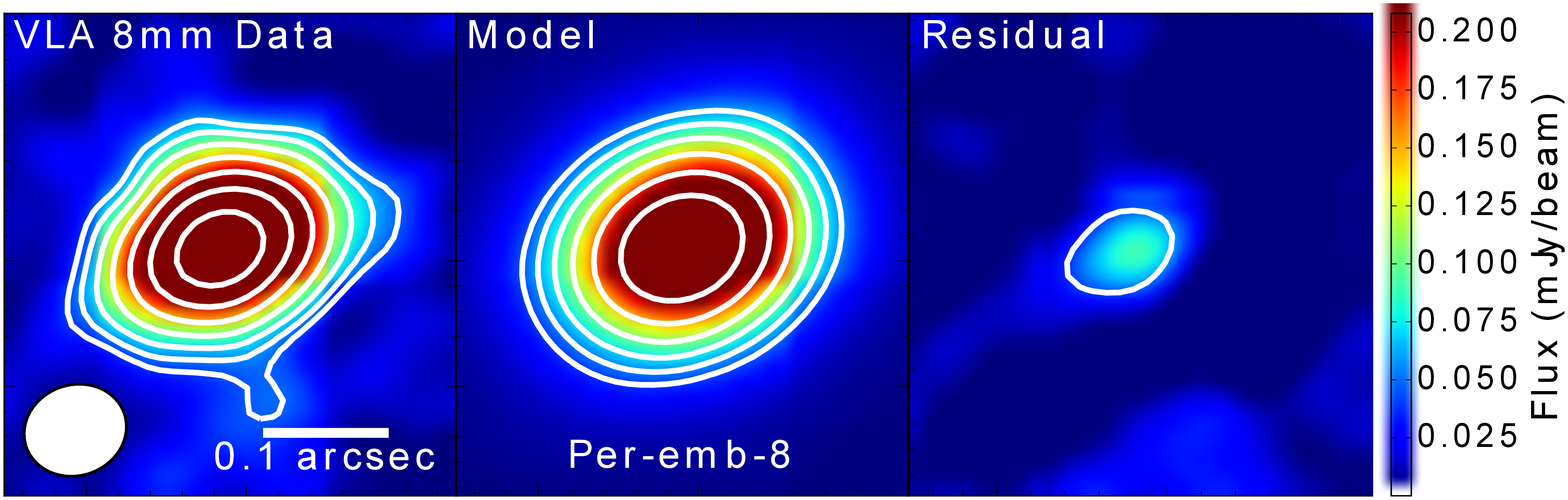}
                \label{fig1g}
        \caption{VLA A+B array data (left), $q=0.25$ model from {\it u,v}-plane best-fit (center), and
residual (right). Images were produced with robust = 0.25 weighting. Contours start at 3$\sigma$ ($\sigma$ $\sim$ 15$\mu$Jy) with a factor of $\sqrt{2}$ spacing. The synthesized beam is in the lower left.
        }\label{fig1}
\end{figure*}

\begin{table*}[t!]
\caption{Observations}
        \vspace{-0.5cm}
\begin{center}
\begin{tabular}{lcccccc}
\hline
Source & $\alpha$ & $\delta$ & A-array  & B-array & Combined Beam & Beam P.A.\\
& (J2000) & (J2000) & Obs. Date  &  Obs. Date & (mas$\times$mas) & ($^{\circ}$)\\
\hline
SVS13B & 03:29:03.078 & +31:15:51.740 & 03 May 2014  & 22 Oct 2013 & 105 $\times$ 83  & -74.8\\
Per-emb-50 & 03:29:07.768 & +31:21:57.125 & 01 Jun 2014 & 26 Oct 2013 &  97 $\times$ 94 & 54.0 \\
Per-emb-14 & 03:29:13.548 & +31:13:58.153 & 25 Feb 2014 & 22 Oct 2013 & 91  $\times$ 75 & 82.4 \\
Per-emb-30 & 03:33:27.303 & +31:07:10.161 &  16 May 2014 & 23 Oct 2013  & 99   $\times$ 92 & -71.5\\
HH211-mms & 03:43:56.805 & +32:00:50.202 & 24 Feb 2014 & 21 Oct 2013 &  96 $\times$ 77 & 85.4\\
IC348 MMS & 03:43:57.064 & +32:03:04.789 & 23 Mar 2014  & 22 Oct 2013  &  89 $\times$ 80 & -57.7 \\
Per-emb-8 & 03:44:43.982 & +32:01:35.209 & 22 Feb 2014 & 01 Oct 2013  & 82 $\times$ 72 & -71.6 \\
\hline
\end{tabular}\\
\end{center}
\tablecomments{Observation dates mark the start of the observations.  Combined beam sizes reflect robust = 0.25 weighting.}
\label{tab1} 
\end{table*}

To characterize the properties of the youngest disks and binaries, we are using  the Karl G. Jansky Very Large Array (VLA) to conduct the VLA Nascent Disk and Multiplicity (VANDAM) continuum survey at $\lambda\sim8, 10 ,40 , 64$ mm toward all identified protostars in the
Perseus molecular cloud \citep{to15a}.
The Perseus molecular cloud is relatively close \citep[d $\sim$ 230 pc;][]{hi08,hi11} and  
 has a significant number of Class 0
and I protostars \citep[43 and 37 systems, respectively;][]{to15p}. 
The early results of the VANDAM survey have already revealed several candidate disk sources 
in both Class 0 and I sources at 10 AU scales.
We consider these to be candidate
disks because we lack kinematic data on small scales to determine whether these structures are rotationally supported.  Nevertheless, so little is known about this early stage of disk evolution that even continuum-only imaging of young disks provides useful constraints on their properties.

In this Letter, we present the first results toward some of the protostellar disk candidates 
around Class 0 and 
Class I sources from the VANDAM survey.  We show the observed structures are 
consistent with disks by fitting 
the 8 mm dust-continuum data in the {\it u,v}-plane to a simple, parameterized
emission model based on the Shakura-Sunyaev disk model.  
VANDAM survey data provide an unparalleled opportunity to study the youngest disks: the unsurpassed  resolution, sample size, and sensitivity for Class 0 and I protostars, permit a detailed examination of typical continuum properties of young disks.\\

 \begin{table*}[t!]
\caption{Source Data}
        \vspace{-0.5cm}
\begin{center}
\begin{tabular}{lccccccc}
\hline
Source & Class  & Deconvolved Size & Disk P.A. & Disk Inclination  & $F_{8mm}$ & $M_{d}$ & $F_{>1700k\lambda}$\\
&   &  (mas$\times$mas) & ($^{\circ}$) & ($^{\circ}$)  & ($\mu$Jy) & (M$_{\odot}$) & ($\mu$Jy)\\
\hline
SVS13B &  0 & 163$\times$80 & 71.4$\pm$2.5 & 61 & 1352.7$\pm$11.6 & 0.14 - 0.29 & 82.0 \\ 
Per-emb-50 &  I &  137$\times$53 & 170.0$\pm$0.3  & 67 &1664.9$\pm$12.5 &  0.18 - 0.36 & 133.2 \\ 
Per-emb-14 &  0 &  174$\times$76 & 12.7$\pm$0.9  & 64 & 882.1$\pm$13.0 & 0.09 - 0.19 & 68.8\\ 
Per-emb-30 &  0 &  87$\times$74 & 40.0$\pm$23.0  &31 & 957.0$\pm$9.4 &  0.10 - 0.21 & 130.9 \\ 
HH211-mms &  0 &  93$\times$59 & 34.8$\pm$9.6  & 51& 867.5$\pm$8.1 &  0.09 - 0.19 & 42.9 \\ 
IC348 MMS & 0 &  145$\times$105 & 70.8$\pm$2.2  & 44 & 1126.5$\pm$10.3 & 0.12 - 0.24 & 0.0\\ 
Per-emb-8 &  0 &  111$\times$84 & 116.1$\pm$2.8  & 41 & 1120.7$\pm$10.3 &  0.12 - 0.24 & 126.5\\ 
\hline
\end{tabular}\\
\end{center}
\tablecomments{Sizes and angles are measured from image-plane 2D Gaussian fits.  Angles are measured counterclockwise from north.  Uncertainties on the deconvolved sizes are $\sim$5.0 mas. Uncertainties on inclinations are $\sim$10$^{\circ}$.}
\label{tab2} 
\end{table*}

\section{Observations}

The VANDAM survey includes Ka-band
 lower-resolution ($\sim$0.28$^{\prime\prime}$ or 65 AU) B-array data
 and high-resolution ($\sim$0.05$^{\prime\prime}$ or 12 AU) 
A-array data. 
We detected 16 protostellar candidate disk sources in the
 Perseus molecular cloud with the data collected in 2013, 2014, and 2015
(Segura-Cox et al. in prep).
 Here we focus on seven of the candidate disks (Table 1; Figure 1, left) which were
observed with A-array prior to 2015 and can be well modeled with the prescription described in Section 4.
 The observations were made in three-bit correlator mode, with a bandwidth of 8 GHz divided into 64 sub-bands. 
 Each sub-band has 128 MHz bandwidth, 2 MHz channels, and full polarization products.  
 The two 4 GHz basebands are centered at 36.9 GHz ($\sim$8.1~mm) and 29.0 GHz ($\sim$10.5~mm).  
 Three sources were observed in each 3.5 hour block, and two sources were observed in each 2.75 hour block.  Some observations were conducted as 1.5 hour blocks.
  3C48 served as the flux calibrator, and 3C84 was the bandpass calibrator. 
 The observations were taken in fast-switching mode to account for rapid atmospheric phase variations, 
 with a 2.5 minute total cycle time to switch between the target source and the complex gain calibrator, 
 J0336+3218.  The total integration time on each source was $\sim$30 minutes for both A-array and B-array.  
 We reduced the data with CASA 4.1.0 and the VLA pipeline (version 1.2.2). 
 We executed additional flagging beyond pipeline flagging by inspecting the phase, gain and bandpass
  calibration solutions.  VLA Ka-band data sets have an estimated amplitude 
  calibration uncertainty of $\sim$10\%, but only statistical uncertainties are considered in our 
  analysis.  \\

 \begin{figure*}[t!]
        \centering
                \includegraphics[width=1.0\textwidth]{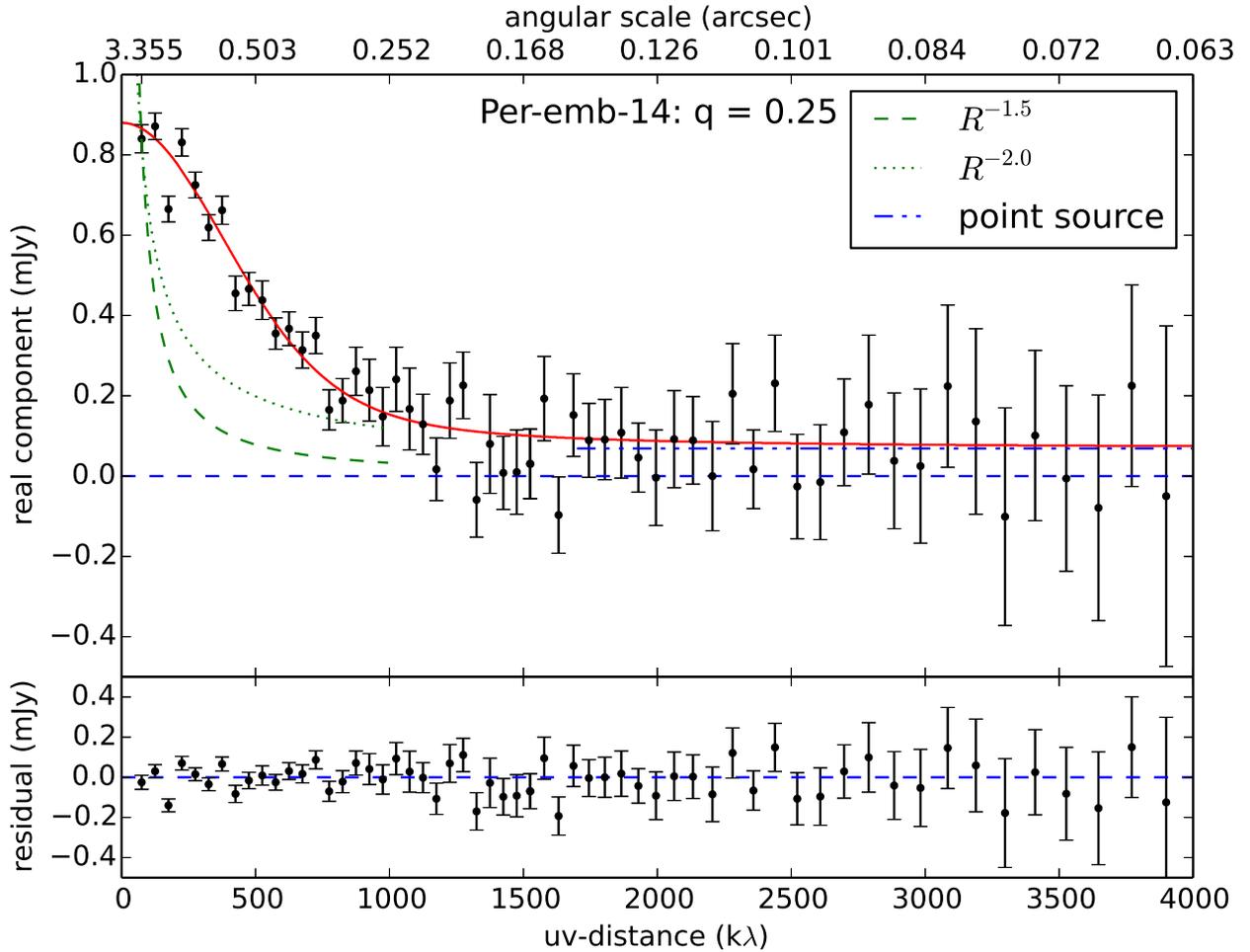}
                \label{fig1a}               
        \caption{Sample real vs {\it u,v}-distance plot of 8 mm data. Top:  real component of data. The blue dashed line indicates real component of zero.  The red solid line is the best-fit
        model.    Bottom: residual of real component minus model.
        }\label{fig2}  
\end{figure*}

\section{Estimated Masses}

Two key quantities that describe disks are their masses and radii.  While accurately determining disk radii requires modeling, 
 masses are estimated from flux measurements (Table 2).
By assuming optically thin emission, we estimate disk masses from the 8~mm dust continuum flux with the relation \citep{hl83}:
\begin{equation}
M=\frac{d^{2}F_{\nu}}{B(T_{d})\kappa_{\nu}},
\end{equation}
where  $F_{\nu}$, $d$, $\kappa_{\nu}$, and $B_{\nu}(T_d)$, are respectively the total observed flux, distance, grain opacity,
and blackbody intensity at dust temperature $T_d$.
We estimate $\kappa_{\nu}$ at 8.1 mm by normalizing to \citet{oh94} at 1.3 mm using a dust to gas ratio of $1/100$: 
$\kappa_{\nu}$ = $(1/100)(\nu/231\mathrm{GHz}))^{\beta}$ cm$^{2}$~g$^{-1}$. $\beta$ = 1 is typically assumed for protostellar disks \citep{an09}, giving $\kappa_{\nu}=0.00146$ cm$^{2}$~g$^{-1}$.
 Mass estimates are inherently ambiguous within an order of magnitude due to uncertainties in the dust-to-gas ratio, $T_d$, and $\beta$; 
rather than compute a single mass estimate, we calculate lower and upper mass estimates for each source
by varying $T_d$. 
Lower masses were estimated by assuming $T_d$ = 40 K, and we determined upper estimates using $T_d$ = 20 K. 
These candidate disks range in flux from 867.5 $\mu$Jy to 1664.9 $\mu$Jy, providing 
 estimated masses of 0.09--0.36 \Ms~(Table 2).\\

\section{Modeling}
We modeled the 8~mm A+B array continuum data  by fitting a symmetric disk intensity profile to the continuum emission of each source.
In this modeling, we deprojected the 8 mm 
visibility data to a fixed position angle (P.A.)
and fixed inclination as determined through image-plane 2D Gaussian 
fitting of the disk candidates and assumed the disks are circularly symmetric (Table 2). We  
azimuthally averaged the visibility data in the {\it u,v}-plane and binned the data in linearly spaced bins of width 50 k$\lambda$  
from 0 to 1500 k$\lambda$, switching to 30 log-spaced bins from 1500 to 4000 k$\lambda$ 
where the data becomes noisier in order to boost the signal-to-noise level at large {\it u,v}-distances.
We accounted for a lower-limit free-free point source component emanating from shocks in the protostellar jets \citep{an98}.  Because point sources in the image domain have constant flux density at all {\it u,v}-distances, we account for the free-free component by calculating the average real component of the binned data with {\it u,v}-distance 
$>$1700 k$\lambda$ and include the average as a flat, linear component of the model (Table 2); the profiles become flat at values $>$1700 k$\lambda$ in all seven sources (e.g., Figure 2).  

We fitted the real components of the deprojected, averaged, and binned profile to a simple
disk model using a C-based implementation of \texttt{emcee}, an affine-invariant Markov chain Monte Carlo ensemble sampler \citep{gw10,fm13}. The imaginary components were assumed to be zero in the model
since the disk is positioned at the phase center and assumed to be symmetric.
 The model mimics  a Shakura-Sunyaev 
disk \citep{ss73} with a power law temperature profile; the resultant model disk surface brightness profile is
\begin{equation}
I(r)\propto\left(\frac{r}{R_{c}}\right)^{-\left(\gamma+q\right)}\exp{\bigg\{\left(\frac{r}{R_{c}}\right)^{\left(2-\gamma\right)}}\bigg\},
\end{equation} 
where $I(r)$ is the radial 
surface brightness distribution, $q$ is the temperature exponent, $\gamma$ is the power-law of the inner-disk surface density, and
$r$ is radius.  $R_{c}$ is a characteristic radius at which there is significant
 optical depth and disk material, making it a proxy for outer disk radius. 
This flux density profile is appropriate here because 
the 8~mm data is expected to be in the Rayleigh-Jeans tail of the dust emission.  
Flux is given by $F = \int_{0}^{\infty} I(r)~2\pi r~dr$.
Flux, disk radius, and gradient were free parameters in the fitting.  We fitted a model disk for values of 
$q=[0.25,0.50,0.75,1.00]$ to the observations in order to avoid over-fitting the data
while exploring reasonable physical values of $q$. \\

\section{Results}

Once model fitting was completed on the candidate disks, 
we generated disk model visibilities corresponding to the best-fit parameters.
The best-fit model and residuals for each source were Fourier transformed, 
  sampled at the same {\it u,v}-points as the data, and imaged using the 
 same weighting as the data (Figure 1, center). 
 We subtracted the model visibility from the data to produce a residual visibility set.  
  Synthetic maps were created of the best-fit model disks. Residual maps were made by imaging the Fourier transformed visibilities after subtracting the model from the data in the {\it u,v}-plane (Figure 1, right).  
An example of the binned observational data, model fits, and residuals is also shown in Figure 2.

Fit results are shown in Table 3.  The $\chi^{2}_{\mathrm{reduced}}$ values near one and the nearly empty residual maps indicate the 
 seven candidate disk sources were well-modeled by our simple-shaped disk profile and are hence likely 
 to be Class 0 and young Class I disks rather than inner envelope structure \citep{ch08}.  
 All candidate disks have modeled $R_{c}$ $>$ 10 AU, and except for HH211-mms the disks are large
 compared to the $R$ = 10 AU upper limit predicted by magnetic braking models of \citet{db10} at the Class 0 stage.

The model includes a disk and point-source component but does not take into account envelope emission.  
The observations filter out the majority of the contaminating envelope emission, demonstrated by 
the visibility profiles corresponding to $R^{-1.5}$ and $R^{-2.0}$ envelope surface density profiles 
plotted in Figure 2.  These envelope density profiles represent, respectively, free-fall collapse and the singular 
isothermal sphere \citep{sh77}. 
The best-fit models were determined from the maximum likelihood, i.e.~the lowest $\chi^{2}$ value, of 
all models fit to the data.  Values of $q$ near 0.5 are predicted by theory \citep[e.g.,][]{cg97} for Class II protostars, which are more evolved than the protostars in our sample.
Lower values of $q$ result in lower values of $\chi^{2}_{\mathrm{reduced}}$ for six candidate disks.  Values of $q$ $<$ 0.5 for Class 0 and younger Class I sources are  favored since the envelope mass reservoir is still large, and radiation is reprocessed from the protostar and directed back onto the disk, increasing the brightness of the outer disk relative to the inner regions of the disk and flattening the brightness distribution  \citep{da96}.  Compared to the other modeled protostars,
$\chi^{2}_{\mathrm{reduced}}$ is notably higher for SVS13B
which may be a more complicated source and not well described by the model. \\

\begin{table}[t!]
\caption{8mm Best-fit Modeling Results}
        \vspace{-0.5cm}
\begin{center}
\begin{tabular}{lcccc}
\hline
Source & $q$ & $\gamma$ & $R_{c}$  & $\chi^{2}_{\mathrm{reduced}}$  \\
&  &  & (AU)  &   \\
\hline
SVS13B & 0.25 & 0.21$^{+0.23}_{-0.20}$ & 24.28$_{-1.7}^{+2.1}$ & 2.194 \\
 & 0.50& 0.42$^{+0.25}_{-0.21}$  &  25.50$_{-1.5}^{+1.9}$  & 2.185  \\
 & 0.75& 0.63$^{+0.24}_{-0.22}$ & 26.46$_{-1.4}^{+1.6}$  & 2.175  \\
 & 1.00& 0.85$^{+0.26}_{-0.23}$ & 27.28$_{-1.2}^{+1.4}$  & 2.164  \\
\cline{1-5}
Per-emb-50 & 0.25& 0.08$^{+0.02}_{-0.16}$ & 21.9$_{-0.9}^{+0.8}$ & 1.556  \\
 & 0.50&  0.26$^{+0.15}_{-0.17}$ & 23.3$_{-1.0}^{+1.1}$ & 1.558 \\
 & 0.75& 0.44$^{+0.16}_{-0.17}$  & 24.6$_{-1.1}^{+1.4}$ & 1.560 \\
 & 1.00&  0.64$^{+0.16}_{-0.18}$ & 25.7$_{-1.3}^{+1.4}$& 1.563 \\
\cline{1-5}
Per-emb-14 & 0.25& -0.11$^{+0.16}_{-0.00}$ &28.5$_{-2.1}^{+2.3}$ & 1.110  \\
  & 0.50& 0.09$^{+0.08}_{-0.21}$ &  30.6$_{-2.3}^{+2.8}$& 1.114  \\
  & 0.75& 0.27$^{+0.17}_{-0.24}$ & 32.5$_{-2.8}^{+2.2}$ & 1.119  \\
  & 1.00& 0.48$^{+0.19}_{-0.23}$ & 33.9$_{-3.1}^{+3.6}$ & 1.123  \\
\cline{1-5}
Per-emb-30 & 0.25&0.02$^{+0.18}_{-0.31}$  & 14.0$_{-0.9}^{+1.0}$& 1.100 \\
 & 0.50& 0.20$^{+0.04}_{-0.32}$  & 14.9$_{-1.1}^{+1.9}$ &1.102  \\
 & 0.75& 0.39$^{+0.30}_{-0.34}$  & 15.8$_{-1.3}^{+1.9}$  &1.104  \\
 & 1.00&0.59$^{+0.14}_{-0.33}$  & 16.5$_{-1.6}^{+31.3}$ &1.107  \\
\cline{1-5}
HH211-mms & 0.25& 0.48$^{+0.40}_{-0.78}$ & 10.5$_{-0.8}^{+0.8}$& 1.009  \\
 & 0.50& 0.65$^{+0.43}_{-0.82}$  & 11.0$_{-0.9}^{+1.0}$ & 1.009  \\
 & 0.75& 0.81$^{+0.42}_{-0.79}$ &11.5$_{-1.2}^{+1.2}$ & 1.009  \\
 & 1.00& 1.01$^{+0.44}_{-0.81}$  &11.9$_{-1.3}^{+1.4}$ & 1.009  \\
\cline{1-5}
IC348 MMS & 0.25& -0.58$^{+0.11}_{-0.11}$ & 25.7$_{-2.2}^{+2.8}$ & 1.085  \\
  & 0.50& -0.39$^{+0.19}_{-0.11}$ & 29.0$_{-2.6}^{+3.2}$& 1.096  \\
  & 0.75&  -0.19$^{+0.11}_{-0.27}$ & 31.6$_{-2.9}^{+4.1}$  & 1.107  \\
  & 1.00& 0.02$^{+0.07}_{-0.11}$ & 33.7$_{-3.1}^{+4.3}$  & 1.118  \\
\cline{1-5}
Per-emb-8 & 0.25& 0.01$^{+0.16}_{-0.19}$  &19.0$_{-1.1}^{+1.2}$ & 1.099  \\
 & 0.50&   0.20$^{+0.17}_{-0.20}$   & 20.2$_{-1.3}^{+1.4}$&  1.107  \\
 & 0.75&   0.40$^{+0.17}_{-0.21}$  &21.2$_{-1.4}^{+1.6}$  & 1.114  \\
 & 1.00& 0.61$^{+0.17}_{-0.20}$ & 22.1$_{-1.6}^{+1.8}$ & 1.122  \\
\hline
\end{tabular}\\
\end{center}
\tablecomments{Values of $q$ are fixed.  Values of $\gamma$ and $R_{c}$ are determined from best-fit models.  Uncertainties reflect 90\% confidence intervals.  }
\label{tab3} 
\end{table}

\section{Discussion}
The estimated masses of the candidate disks are consistent with known disks around Class 0 and I protostars. 
The seven VANDAM candidate Class 0 and I disks range in estimated masses of 0.09--0.36 \Ms. 
Scaling to our value of $\kappa_{\nu}$, Class 0 protostar L1527's disk is 0.013 \Ms~\citep{to13} with $T_d$ = 30 K. 
A recent interferometric study of Class I disks in Taurus \citep{ha14} revealed the disks around TMC1A, 
TMC1, TMR1 and L1536 to have masses of 0.4--3.3$\times$10$^{-2}$ \Ms~with $T_d$ = 30 K and \citet{oh94} opacities.
All mass estimates have uncertainties of a factor of $\sim$10 due to the unknown gas-to-dust ratio, $T_d$, $\beta$, and varying choices of $\kappa_{\nu}$.

Values of $\gamma$, the power-law of the inner-disk surface density, for the VANDAM sources 
can be compared to values found for older Class II sources.   The best-fit models for our seven sources yield $-0.58<\gamma<0.45$. Negative values of $\gamma$ indicate increasing surface density with radius, inconsistent with disks, yet for all sources at least one value of $q$ exists which produces a positive best-fit $\gamma$ (Table 3).  \citet{an10} determined $\gamma$ at 880 $\mu$m for disks around Class II 
objects using a two-dimensional parametric model with a prescription of the surface density profile similar to
 the radial surface brightness distribution applied in this Letter.   The Class II sources have values of 
$0.4<\gamma<1.1$, larger than VANDAM results.  
 The modeled, surface density profiles of the Class 0 and I systems  
 taper off less quickly with radius compared to the Class II systems inside the characteristic radius $R_c$.

The best-fit model radii of 
the candidate Class 0  disks are between $\sim$15-30 AU, except HH211-mms with $R_{c}$ $\sim$ 10 AU.   These are smaller 
than the Keplerian disks found in L1527 and VLA 1623 at 1.3 mm
  \citep[$R$ $\sim$ 54 AU and $R$ $\sim$ 189 AU respectively]{oh14,mu13} 
  but consistent with the size of HH212  \citep[R $>$ 30 AU]{cd14}.  The modeled radii (Table 3) are a factor of 1 to 1.5 times larger than the deconvolved sizes (Table 2). 
  The radii of the new candidate
disks may represent lower limits on disk sizes because continuum emission from dust is biased by dust
grain size.  Radial drift \citep{bi10,we77} sends large grains inward in the disk, and so the long wavelength observations
 preferentially trace
inner disk emission \citep{pe12}.  We expect the 8 mm dust-modeled radii to be less than the gas disk radius, and so the 8 mm 
radii are lower-limits on disk size; multi-wavelength observations that include gas tracers are required to further constrain disk radii.
Per-emb-14 was resolved at 1.3 mm by \citet{to15b} with $R$ $\sim$ 100 AU; thus the disk is indeed likely larger than the 8 mm radius quoted here.
 In all cases, the previously known disks and new candidate Class 0 disks are larger than the expected 
 upper limit of 10 AU from  strong magnetic braking models \citep[e.g.,][]{db10}.

The seven VANDAM candidate disks are well-fit by a disk-shaped model and have disk
masses, values of $\gamma$, and radii consistent with known disks.  These candidate disks can be
compared to the well-studied Class 0 systems with disks: L1527 \citep{to12} and VLA 1623 
\citep{ml13}.  L1527 and VLA 1623 have large Keplerian disks \citep[$R$ $\sim$ 50 AU,][]{oh14,mu13}  as well as misaligned magnetic fields and rotation axes \citep{hu14}, suggesting that misaligned fields and large disks may be linked \citep{sc15}.   
 The discrepancy between theoretical and observed disk sizes could be due to misalignment between the magnetic field and rotation axis, which changes the strength of magnetic braking, allowing disks to grow at early times  \citep{hc09,jo12,li13,ku13}.  
 Similarly, aligned orientations would strengthen magnetic braking and limit disk size, as seen in Class 0 source B335 with a disk of $R$ $<$ 5 AU \citep{ye15,hu14}.
The large Keplerian Class 0 disks and new  candidate disks indicate significant magnetic braking either has already occurred, has not happened, or the magnetic field is weak enough for disks with $R$ $>$ 10 AU around these Class 0 protostars to form, though other configurations are possible.

It is illustrative to compare the presence of VANDAM candidate disks to the magnetic field morphology of the systems. The TADPOL survey examines protostellar dust polarization and hence plane-of-sky magnetic field orientation. \citep{hu14}.  For sources with detected magnetic fields,
 we are able to determine
 which systems have misaligned 
 orientations between the outflow  (a proxy for rotation axis) and the average 
 magnetic field orientation.
The magnetic field morphologies of two of the seven VANDAM sources in this Letter (HH211-mms and SVS13B) were also examined as a part of the TADPOL survey.  
  A qualitative 
comparison between the VANDAM data and the TADPOL magnetic field orientation reveals that   
HH211-mms and SVS13B morphologies agree with
the tentative link between $R$ $>$ 10 AU disks and misaligned orientations \citep{sc15}. 
HH211-mms has a clearly misaligned orientation \citep{hu14}, agreeing with theoretical predictions that disks with $R$ $>$ 10 AU disks form at early times in systems with non-parallel orientations.
SVS13B has a more complicated magnetic field morphology with an average aligned orientation \citep{hu14} with the outflow \citep{ba98}, however the TADPOL data has lower resolution than our data ($\sim$3$^{\prime\prime}$, or 700 AU), and the innermost regions of the magnetic field has misaligned orientation near the candidate disk.  The small-scale, misaligned orientation in SVS13B could be caused by magnetic fields being wrapped up with the rotating disk, and the aligned magnetic field lines further from the disk are  less influenced by the disk wrapping.\\

\section{Conclusions}
The first results of the VANDAM survey revealed new candidate Class 0 and I disks in continuum 
emission in the Perseus molecular cloud and is the most sensitive, most complete, and highest resolution survey of young protostellar disks to-date.  
 For seven of the VANDAM systems, we fit the deprojected, averaged, and binned 
data in the {\it u,v}-plane to a disk-shaped profile, accounting for a free-free point source potentially contaminating the dust emission, 
to affirm the disk candidacy of the sources and to begin to model disk properties.  To confirm that these disks are true, rotationally-supported disks, kinematic follow-up observations are
required.   The seven VANDAM candidate disks are well-fit by a model with a disk-shaped profile and have masses comparable to known disks. The inner-disk surface densities of the Class 0 and I candidate disks taper off less quickly with radius than their Class II counterparts. 

 The best-fit model radii of the new candidate disks are $R_{c}$ $>$ 15 AU, except HH211-mms with $R_{c}$ $\sim$ 10 AU.
Since magnetic braking is expected to suppress Class 0 disks to $R$ $<$ 10 AU, this
suggests that magnetic braking has already ended or is inconsequential in some Class 0 sources.  At
8 mm, the modeled radii are lower limits on the disk size since dust continuum emission is tied to grain
size and large grains radially drift inwards in the disk; the actual sizes of the disks may be larger.  
Multi-wavelength observations are needed to put added constraints on disk radii.  Furthermore, 
theory predicts larger disk growth in systems with misaligned outflow axes and magnetic field 
orientations, and the early VANDAM candidate disk results combined with polarization data for two of 
the sources indicate that indeed $R$ $>$ 10 AU disks have formed in systems with misaligned 
orientations.


\acknowledgments
This research made use of APLpy, an open-source plotting package for Python hosted at http://aplpy.github.com. 

DMSC is currently supported by  NRAO Student Observing Support grant SBC NRAO 2015-06997.  JJT is currently supported by grant 639.041.439 from the Netherlands
Organisation for Scientific Research (NWO). ZYL is supported by NASA NNX14AB38G and NSF AST-1313083.

 The National Radio Astronomy
Observatory is a facility of the National Science Foundation
operated under cooperative agreement by Associated Universities, Inc.
\\
\\


\clearpage

\end{document}